\documentclass[a4paper,11pt]{article}

\usepackage{jinstpub} 

\usepackage[T1]{fontenc}         
\usepackage[]{graphicx}          
\usepackage{subcaption}          
\graphicspath{{./}}

\usepackage[german,english]{babel}   
\usepackage{microtype}                
\usepackage{hyperref}                 
\usepackage{cleveref}                 
\usepackage{eurosym}                  
\usepackage{siunitx}                  
\usepackage[version=4]{mhchem}        
\usepackage{amsfonts,stackrel}        
\usepackage{pgf}                      
\usepackage{layouts}                  

\usepackage{tikz}
\usetikzlibrary{shapes.geometric,arrows,positioning}
\usetikzlibrary{shadings,fadings,backgrounds,fit}

\usepackage{shellesc}                 
\usetikzlibrary{external}             
\definecolor{ippBlue}{RGB}{0, 101, 191}
\definecolor{ccfeOrange}{RGB}{243, 148, 0}
\definecolor{mpGreen}{RGB}{0, 123, 108}
\definecolor{taRed}{RGB}{217,41,59}
\definecolor{duPurple}{RGB}{97,28,102}

\hypersetup{
  colorlinks=true,               
  linkcolor=black,               
  filecolor=ccfeOrange,          
  urlcolor=ippBlue,              
  citecolor=ippBlue,              
  pdftitle={ICFRD 2020 : DEMO interferometer/polarimeter simulations }, 
}

\tikzset{
  treenode/.style   = {shape=rectangle,
    draw=black, very thick, text=white, fill=ippBlue,
    inner sep=1mm, minimum height = 15mm},
  optical/.style   = {shape=rectangle,
    draw=black, fill = white, very thick,
    inner sep=1mm},
  SToptical/.style   = {optical,
    fill opacity=0.7, text opacity = 1},
  STznse/.style   = {SToptical, fill = mpGreen},
  STaggase/.style   = {SToptical, fill = black, fill opacity=0.5},
}

\pgfdeclareradialshading{plasma}{\pgfpoint{0cm}{0cm}}%
{
  color(0cm)=(ippBlue!50);
  color(3mm)=(ippBlue!50);
  color(7mm)=(taRed);
  color(10mm)=(ippBlue!50)
}

\newcommand{\includepdf}[1]{\includegraphics[]{./#1.pdf}}

\newcommand{\myvec}[1]{\ensuremath{\begin{pmatrix}#1\end{pmatrix}}}

\newcommand{\ipp}{Max-Planck-Institute for Plasma Physics, Wendelsteinstr. 1, 17489 Greifswald, Germany}

\title{Design considerations of the European DEMO’s IR-interferometer/polarimeter based on TRAVIS simulations}

\author[a,1]{K. J. Brunner,\note{Corresponding author.}}
\author[a]{N. Marushchenko,}
\author[a]{Y. Turkin,}
\author[b]{W. Biel,}
\author[a]{J. Knauer,}
\author[a]{M. Hirsch,}
\author[a]{R. Wolf}

\affiliation[a]{\ipp}
\affiliation[b]{Forschungszentrum Jülich, Wilhelm-Johnen-Straße, 52428 Jülich, Germany}

\emailAdd{k.j.brunner@ipp.mpg.de}

\abstract{Interferometry is the primary density control diagnostic for large-scale fusion devices, including ITER and DEMO. In this paper we present a ray tracing simulation based on TRAVIS accounting for relativistic effects. The study shows that measurements will over-estimate the plasma density by as much as \SI{20}{\degree}. In addition, we present a measurement geometry, which will enable vertical position control during the plasma's ramp-up phase when gap-reflectometers and neutron cameras are still blind.}


\arxivnumber{2112.09363} 

\def\acknowledgement{\itshape This work has been carried out within the framework of the EUROfusion Consortium, funded by the European Union via the Euratom Research and Training Programme (Grant Agreement No 101052200 — EUROfusion). Views and opinions expressed are however those of the author(s) only and do not necessarily reflect those of the European Union or the European Commission. Neither the European Union nor the European Commission can be held responsible for them.}

\usepackage{layouts}
\begin{document}

\maketitle
\flushbottom 


\section{Introduction}
\label{sec:intro}

Plasma density control, which will be mandatory on the European demonstration reactor (DEMO), will not only require actuators such as gas puffing and pellet injection, but also a reliable measurement of the plasma density\cite{Biel2019}. There are robust density diagnostics foreseen, such as reflectometry and neutron cameras\cite{Malaquias2018,Biel2019}. However, these systems are unreliable during the early phases of the plasma when the plasma is not yet fully formed and no neutrons are generated. An additional density control diagnostic during these operation-periods is therefore necessary. Further difficulties arise from the fact that a Tokamak-based DEMO will be experience instabilities of the vertical plasma position. During the developed plasma phase gap-reflectometers are able to keep the plasma in its design position. However, during the ramp-up phase they are blind, since the plasma is small and too far away. In consequence the curvature of the cut-off surface is too large so that the reflected probing beam diffuses too much to be useful. Magnetic diagnostics are too slow in DEMO to make up for this gap and hence another diagnostic is required to constrain the vertical plasma position during the early ramp-up phase.

Interferometry is one of the most robust density diagnostics on large-scale fusion experiments today and is therefore foreseen for both ITER and DEMO\cite{Boboc2012,Sasao2016,vanZeeland2017,Brunner2018,,Biel2019}. However, the inherent 2$\pi$ phase ambiguity makes interferometry a questionable choice for a robust high-plasma-density control diagnostic. To resolve this, a polarimeter can be added to the optical design\cite{vanZeeland2017,Akiyama2016b}. Just like any other diagnostic-system it comes with a set of particular challenges when integrating it into the DEMO device. One of them is the effect of relativistic corrections to the dispersion relation.

In this paper we present an investigation into these effects using the current geometry envisaged for the DEMO multi-channel interferometer/polarimeter~(MCIP) device. We use the TRAVIS ray tracer and a polarization tracer based on the Stokes formalism\cite{Marushchenko2014}. Simulations are based on plasma parameters and equilibria from the 2019 DEMO baseline. Similar approaches to model interferometric phase shift were previously conducted for NSTX\cite{Zhang2010}. Using the simulations we propose a measurement geometry for the MCIP system to provide vertical feed-back information during the plasma's ramp-up phase.

\section{Mathematical Background}
\label{sec:background}

In most machines operating today the interferometric phase shift induced by the plasma can be treated by the cold plasma dispersion :

\begin{equation}
  \label{eq:interfPhase}
  \phi \approx 2.82\times10^{-15} \left[\si{\radian \m}\right] \times \lambda \int n_\text{e} \text{d}z.
\end{equation}

Here $\lambda$ is the laser wavelength and $n_\text{e}$ is the plasma electron density integrated along the propagation path $z$. The evolution of the polarization ellipse can be described via the M\"{u}ller-Stokes-formalism\cite{Segre1999}:

\begin{equation}
  \label{eq:stokesEqn}
  \frac{\text{d}\boldsymbol{s}}{\text{dz}} = \boldsymbol{\Omega}\times \boldsymbol{s},
\end{equation}

with

\begin{equation}
  \label{eq:stokesVec}
  \myvec{s_1\\s_2\\s_3} = \myvec{\cos{2\chi}\cos{2\psi}\\\cos{2\chi}\sin{2\psi}\\\sin{2\chi}}
\end{equation}

\begin{equation}
  \label{eq:coldOmega}
  \boldsymbol{\Omega}_\text{cold} = \frac{\omega_\text{pe}^2 \omega_\text{ce}^2}{2 c \omega \left( \omega^2 -  \omega_\text{ce}^2 \right)} %
  \myvec{\left(B_x^2 - B_y^2\right)/B^2\\2 B_xB_y/B^2\\2(\omega/\omega_\text{ce})B_z/B}.
\end{equation}

In these equations $\boldsymbol{s}$ denotes the reduced Stokes-vector, where we assume that the probing beam is constant in amplitude thus eliminating one component\cite[eq.(1)]{Zhang2010}. Its evolution along the path of propagation $z$ is described by the vector of angular velocity $\boldsymbol{\Omega}$, which is defined by the \emph{local} plasma parameters. In \cref{eq:coldOmega} $c$ denotes the vacuum speed of light and $\omega$ the probing beam's angular frequency. $\omega_\text{pe}$ and $\omega_\text{ce}$ are the electron plasma and cyclotron frequency, respectively. $\boldsymbol{B}$ marks the magnetic field vector with the $z$ component parallel to the direction of propagation. The angles $\chi$ and $\psi$ in \cref{eq:stokesVec} are the ellipticity and polarization angle of $\boldsymbol{s}$, respectively. The evolution of $\chi$ and $\psi$ is coupled via \cref{eq:stokesEqn}. More explicitly, $\psi$ changes both due to Cotton-Mouton~(CM) and Faraday rotation~(FR) effect, whereas the evolution of $\chi$ is modified according to the local value of $\psi$\cite{Guenther2004}. Using the full Stokes equation \ref{eq:stokesEqn} the coupling between the two can be accurately modeled.

It was previously established that ITER and DEMO parameters require relativistic corrections to the interferometric phase\cite{Mirnov2018}:

\begin{equation}
  \label{eq:relCorrectionInterf}
  \frac{\Delta\Phi_{\text{interf.}}}{\Phi_{\text{interf.,cold}}} = \left( - \frac{3}{2} \int \frac{n_e T_e}{m_e c^2}\text{d}z + \frac{15}{8} \int \frac{n_e T_e^2}{m_e^2 c^4}\text{d}z \right) / \int n_e \text{d}z.
\end{equation}

Similarly, it was shown that the angular velocity vector $\boldsymbol{\Omega}$ requires relativistic correction modifying the coupling between the FR effect and the CM effect as

\begin{equation}
  \label{eq:relCorrectionOmega}
  \boldsymbol{\Omega} = \boldsymbol{\Omega}_\text{cold} \cdot \left[ 1 + \frac{T_e}{m_e c^2} \myvec{9/2\\9/2\\-2}  + \frac{T_e^2}{m_e^2 c^4} \myvec{15/8\\15/8\\3} \right].
\end{equation}

These corrections are highly relevant for the DEMO parameter space.

\section{Simulation of a ramping DEMO discharge}
\label{sec:simulation}

To investigate their effect the TRAVIS code was used to trace an interferometer ray through the DEMO plasma\cite{Marushchenko2014}. This code is an established ray tracing code primarily intended for microwave based diagnostics operating in the Gigahertz-range. However, the fundamental physics governing propagation do not differ to the infra-red (IR) wavelength range, so that TRAVIS can be used to model an interferometer ray.

\subsection{Simulation setup}
\label{ssec:simBG}

\begin{figure}[b]
  \centering
  \begin{subfigure}{0.45\textwidth}
    \includepdf{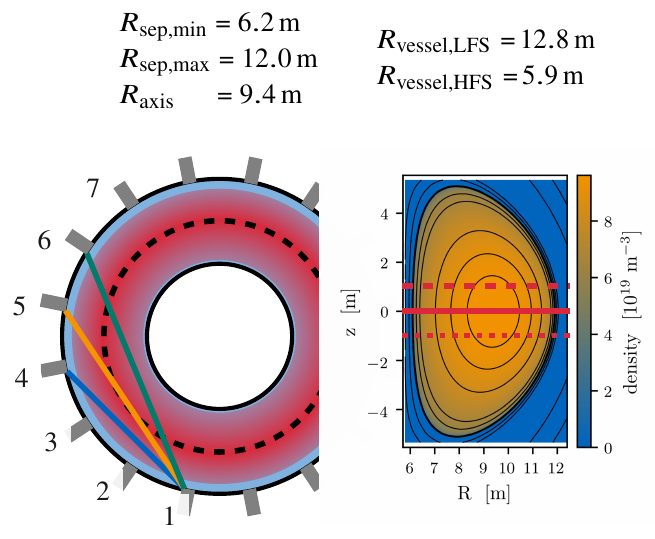}
    \caption{\label{fig:geometry} The geometry of the simulated MCIP. The toroidal geometry is on the left; the poloidal geometry on the right. The colored lines indicate the probing beam paths. The three red beams on the right are at the limit of the port plug space available for a vertical array.}
  \end{subfigure}
  \hfill
  \begin{subfigure}{0.45\textwidth}
    \includepdf{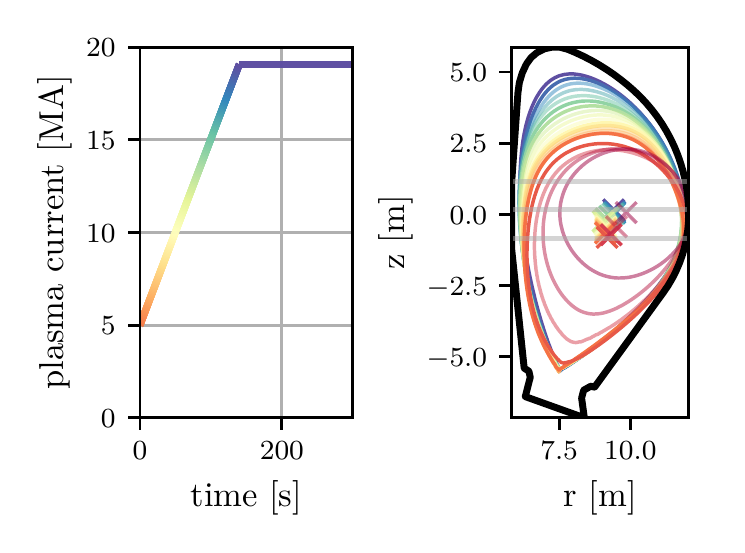}
    \caption{\label{fig:rampOverplot} The plasma parameters used for the simulations. The plasma current used for mapping equilibria to density and temperature profiles are on the left. The correspondingly colored  equilibria used for simulations are on the right.}
  \end{subfigure}
  \caption{\label{fig:simsetup}The simulation setup used for TRAVIS-calculations. The geometry is depicted in (a), the equilibria are depicted in (b).}
\end{figure}

\Cref{fig:simsetup} depicts the relevant input parameters for the TRAVIS simulations. \Cref{fig:geometry} shows the simulation geometry. The MCIP is set up similar to the ITER TIP system with a toroidal geometry leading from one primary launch port to corner-cube reflectors~(CCR) in other ports. As seen on the left, only 3 port combinations sample the plasma : $1\rightarrow4$~(blue), $1\rightarrow5$~(orange) and $1\rightarrow6$~(green). TRAVIS traces the plasma parameters along the ray. All traces are a single pass through the plasma from the launch port 1 to the respective receiving port. Since TRAVIS can only launch pure O/X-mode rays, the ray modelled by it is launched in ordinary-mode polarization. TRAVIS is capable of using the weakly relativistic dispersion relation for polarization and refraction tracing. For refraction the difference between the two is negligible, however for polarization propagation a pure-mode launch was undesirable. As such TRAVIS' primary role is refraction and parameter tracing and handling all the geometry. The trace results are then passed to a separate Python-based tracer, which is used to handle interferometric phase shift and polarization according to \cref{eq:interfPhase,eq:stokesEqn} using the relativistic corrections described by \cref{eq:relCorrectionInterf,eq:relCorrectionOmega}. 

\Cref{fig:rampOverplot} shows the equilibria used for the simulations. The equilibria are based on calculations by Mattei \emph{et al.}\cite{Mattei2019eurofusion}. These in turn are based on plasma parameters from the ``orange case'' scenario by Palermo \emph{et al.}\cite{Palermo2020eurofusion}. The underlying profiles are shown in the bottom of \cref{fig:measTrace}. The two data-sets are matched based on the toroidal current shown in the left of \cref{fig:rampOverplot}. The correspondingly colored equilibria are shown on the right. Note that only the diverted equilibria are part of the simulation.

\subsection{Refraction}
\label{ssec:refraction}

The first simulation's aim was to evaluate, whether refraction limits the wavelength choice for the DEMO MCIP system. For this a simple equilibrium as shown in the right of \cref{fig:geometry} was used and the probing beam shifted up and down with respect to the current centroid. The distance of the ``target position'' when exiting the plasma was tracked by TRAVIS. The wavelength was varied from \SIrange{5}{100}{\um}.

The data -though not depicted here- shows that even in a static scenario the wavelength beyond \SI{11}{\um} is unfeasible for a two-colored interferometer. Such a system (whether implemented as traditional two-color system or dispersion interferometer) is mandatory in large continuously-operating device to compensate vibrations on the measurements. The beam separation of a Methanol-laser based two-color system running at \SIlist{50;44}{\um} would be of the order of \SI{1}{\mm}. At this level of separation, the effectiveness of the vibration compensation would be in jeopardy. As such, only the medium IR wavelength range around \SI{10}{\um} is acceptable with only 10s of \si{\um} in beam separation.

In general, refraction must be minimized. This is primarily, because the DEMO MCIP diagnostic will have to also minimize duct sizes for mirror protection. Currently it is estimated, that only \SI{40}{\mm} ducts can be accepted per measurement beam, i.e. going towards and returning from a CCR in the receiving port. A \SI{1}{\mm} displacement by refraction alone will prove very difficult to compensate by beam steering on the return beam.

\subsection{Diagnostic trace}
\label{ssec:trace}

\begin{figure}
  \centering
  \includegraphics{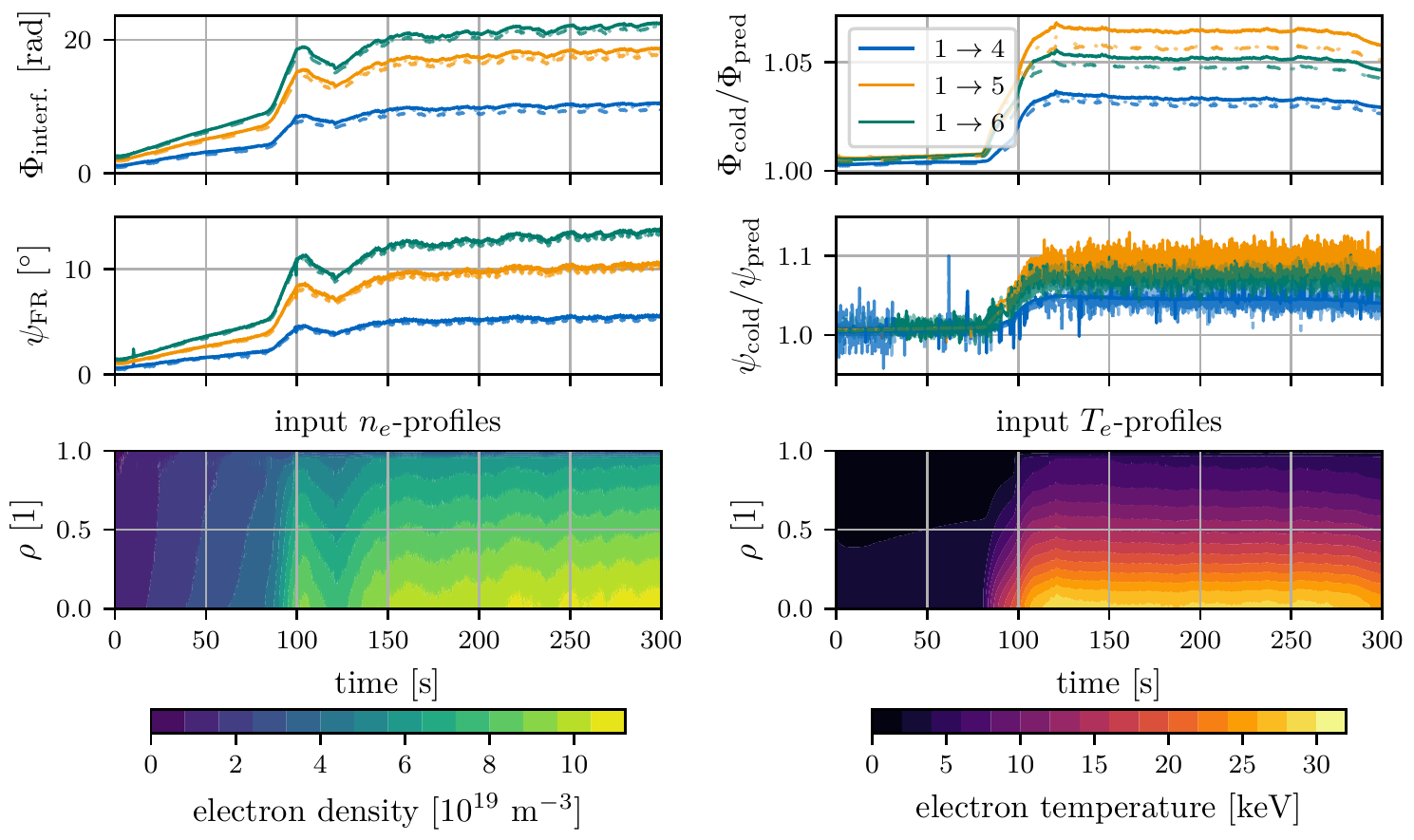}
  \caption{\label{fig:measTrace} The simulated measurements of an interferometer (top left) and a Faraday polarimeter (left center) accounting for relativistic effects. The correction factors to the corresponding cold dispersion are on the right. The bottom two plots show the plasma parameters (density on the left; temperature on the right) from the Palermo \emph{et al.} ``orange case'' scenario\cite{Palermo2020eurofusion}. The propagation path is as indicated in \cref{fig:simsetup}.}
\end{figure}

Following the results of \cref{ssec:refraction} a diagnostic trace was simulated. The colors of the traces correspond to the toroidal beam path seen in \cref{fig:geometry}. \Cref{fig:measTrace} shows the expected interferometric phase-shift (top left) and FR angle (left center) for a \SIlist{10;5}{\um} interferometer/polarimeter passing through the plasma once taking relativistic effects into account. The respective trace on the right shows the relative error one makes, when assuming the cold dispersion.

Apart from showing significant interferometric phase shift (of the order of 3 to 8 $\pi$), a noteworthy result is that both the interferometric phase as well as the FR angle are being over-estimated using the cold plasma dispersion. This means that a failure of the correction would bring a density controller farther away from the Greenwald-density limit and hence into a ``safer'', albeit sub-optimal, operation regime.

The simulation shows in addition that there is a significant difference between the on-axis (solid lines) and off-axis view chords (dotted/dashed lines). For reference the corresponding lines are marked in color on the left in \cref{fig:geometry} and with the corresponding pattern on the right. The difference between the off-axis and on-axis measurement is utilized for the estimation of the current-centroid's vertical position.

\subsection{Vertical stability feedback trace}
\label{ssec:viTrace}

Vertical stability feedback information requires the knowledge of the position of the current centroid. To measure this, one can use the fact that a Tokamak density profile tends to be peaked. Even for DEMO, where transport effect are expected to flatten the profile, it is still expected to quasi-monotonously increase towards the magnetic axis\cite{Fable2019}. As such, it is possible to estimate the position of the current centroid by stacking several interferometric chords above each other. The current centroid will be the position of maximum $\phi$ or $\psi_\text{FR}$, which can be found by a simple nearest neighbor interpolation.

\begin{figure}[t]
  \includegraphics{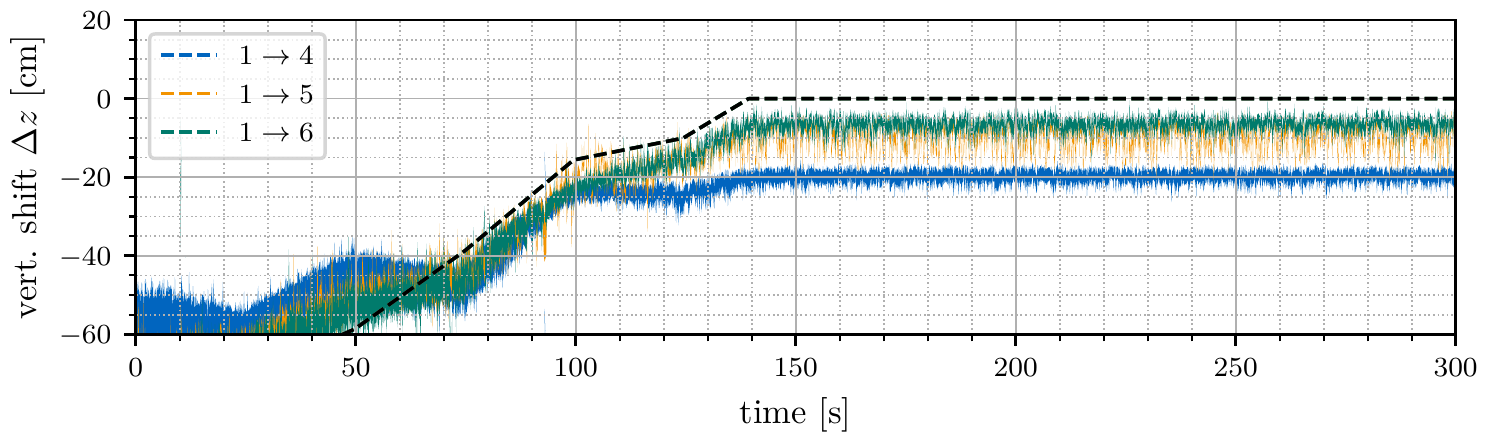}
  \caption{\label{fig:viFBTrace} The vertical feed-back trace calculated from the data in \cref{fig:measTrace} using a 7 chord vertical array. The measurement error is assumed to be \SI{10}{\milli\radian} indicated by the line width. $\Delta z$ is the measured distance to the target. The dashed black line is the true plasma position.}
\end{figure}

To evaluate whether the MCIP system can deliver such feed-back information, the port space between the dashed and dotted line on the right in \cref{fig:geometry} is filled with seven such chords stacked on top of each other resulting in roughly \SI{46}{\cm} vertical spacing between the measurement chords. The use of only three chords was not performing well due to the plasma's shape. The results are shown in \cref{fig:viFBTrace}. The position of the current centroid is estimated based on a linear interpolation between nearest (vertical) neighbors for each of the chord combinations. This will then deliver an estimate for the position of the current centroid and its vertical velocity. While this is not an accurate measurement, it ought to suffice to prevent the plasma from hitting the wall before the gap reflectometers can operate reliably.

The measurement error in the simulation is assumed to be \SI{10}{\milli\radian}, which is achievable with systems such as the ITER TIP\cite{vanZeeland2017}. They are indicated by the line width. For reference the corresponding lines are again marked in color on the left in \cref{fig:geometry}. The measured position to the target (the central measurement chord) $\Delta z$ is on the left. The true plasma position is indicated by the black dashed line.

As can be seen from the simulation, the $1 \rightarrow 6$ port combination performs best. This is the result of the strongest difference between the neighboring measurement channels, which is to be expected considering that this port combination also measures the highest phase shift. The $1 \rightarrow 5$ combination measures only slightly less accurate. The estimated vertical position is wrong by \SIrange{5}{10}{\cm}. The $1 \rightarrow 4$ port combination appears to perform rather poorly, which may again be due to the strong shaping and the fact that this port combination does not pass through the plasma center.

A point of note: there was an error in the axis definition of the equilibria used for the simulation, which could not be corrected. This may account for some of the residual offset seen.

\section{Summary \& Discussion}
\label{sec:discussion}

We have presented the results of a ray tracing analysis for the multi-channel interferometer/polarimeter diagnostic~(MCIP) for the European DEMO based on TRAVIS. A refraction analysis based on these results shows that the diagnostic will have to operate using a \ce{CO2}-based Laser system or something similar operating in the \SI{10}{\um} wavelength range. Higher wavelength lasers will experience unacceptable levels of color separation. The simulations also show that the MCIP diagnostic layout will suffer from relativistic effects with a measurement error of \SIrange{10}{20}{\percent} in a double-pass scenario (the simulations were conducted in single pass). The error on the polarimetric measurement is of similar scale. We have also shown that a vertically stacked array of 7 interferometer chords can estimate the position of the plasma's current centroid during the DEMO ramp-up phase by nearest neighbor interpolation. The simulations will however have to be run again using improved scenarios and equilibria.

A similar approach to vertical stability control has previously been conducted at EAST using polarimetry where the results shown here could be experimentally verified\cite{Lian2019}. In the next iteration of the MCIP diagnostic design the integration of such a vertical array into the limited port space of the DEMO device needs to be evaluated. By geometrically optimizing the probing beam spacing it may be possible to improve the accuracy of the vertical position measurement whilst reducing the number of necessary beams. The tool developed here can be expanded to integrate into a full beam transport model for a synthetic diagnostic. 

\acknowledgments

\acknowledgement

\section*{Sources}

Much of the evaluation in this article was conducted using Python 3.9 in combination with the numpy, matplotlib and pandas libraries\cite{python,pandas,matplotlib,numpy}.

Data supporting the findings of this study as well as the codes for data evaluation and plot generation can be supplied upon reasonable request by sending an application to the corresponding authors at \ipp.


\bibliographystyle{JHEP}
\bibliography{papers}

\end{document}